\documentclass[twocolumn,showpacs,preprintnumbers,amsmath,amssymb]{revtex4-2}

\usepackage{graphicx}
\usepackage{bm}

\begin{document}

\preprint{}

\title{Splitting of the three-body F\"orster resonance in Rb Rydberg atoms \\ as a measure of dipole-dipole interaction strength}

\author{I.~I.~Ryabtsev$^{1, 2}$}
  \email{ryabtsev@isp.nsc.ru}
\author{I.~N.~Ashkarin$^{3}$}
\author{I.~I.~Beterov$^{1, 2, 4}$}
\author{D.~B.~Tretyakov$^{1}$}
\author{E.~A.~Yakshina$^{1, 2, 4}$}
\author{V.~M.~Entin$^{1}$}
\author{P.~Cheinet$^{3}$}

\affiliation{$^1$Rzhanov Institute of Semiconductor Physics SB RAS, 630090 Novosibirsk, Russia }
\affiliation{$^2$Novosibirsk State University, 630090 Novosibirsk, Russia}
\affiliation{$^3$Universit\'e Paris-Saclay, CNRS, Laboratoire Aim\'e Cotton, 91405 Orsay, France}
\affiliation{$^4$Institute of Laser Physics SB RAS, 630090, Novosibirsk, Russia}

\date{\today}

\begin{abstract}

Three-body F\"orster resonances controlled by a dc electric field are of interest for the implementation of three-qubit quantum gates with single atoms in optical traps using their laser excitation into strongly interacting Rydberg states. In our recent theoretical paper [Zh. Eksper. Teor. Fiz. \textbf{168}(1), 14 (2025)], it was found that the proposed earlier three-body F\"orster resonance $3\times nP_{3/2} \to nS_{1/2} +(n+1)S_{1/2} +nP_{1/2} $ in Rb Rydberg atoms presents a splitting, with one of the split components having weaker dependence of the resonant electric field (and the corresponding dynamic shift) on the distance $R$ between the atoms. Here we study this effect in more detail, since this resonance appears as the most suitable for performing experiments on observing coherent oscillations of populations of collective three-body states and implementing three-qubit quantum gates based on them. For a linear spatial configuration of three interacting Rydberg atoms, the physical mechanism of this phenomenon is revealed and analytical formulas are obtained that describe the behavior of split structure of the F\"orster resonance depending on $R$. It is found that the splitting is a measure of the energy of the resonant dipole-dipole exchange interaction with an excitation hopping between neighboring Rydberg states $S$ and $P$.

\end{abstract}

 \maketitle

\section{Introduction}

In recent years, significant progress has been made in implementing quantum computations and simulations with qubit registers based on single neutral atoms in large arrays of optical dipole traps \cite{1,2,3,4,5}. The execution of entangling quantum gates or quantum simulations is achieved by laser excitation of atoms to Rydberg states. For neutral-atom qubits, fidelities of two-qubit gates exceeding 0.994 have been reached \cite{5,6,7}. This fidelity, however, remains insufficient for quantum computation without error correction.

Error correction is performed based on multi-qubit quantum gates, such as the three-qubit Toffoli gate(CCNOT gate) \cite{8,9,10}. The implementation of such a gate requires controlled simultaneous interaction of three Rydberg atoms.
The first experimental implementation of the Toffoli gate based on a dipole blockade effect reported a fidelity of 87\% \cite{11}. The use of dipole blockade requires strong interactions of Rydberg atoms \cite{9,12,13}, so further increase in the fidelity of three-qubit gates based on it remains questionable.

In our theoretical papers \cite{14,15,16} it was proposed to use coherent oscillations of populations and phases of three-body collective states at three-body F\"orster resonances controlled by a weak electric field to implement the Toffoli gate. This method does not require strong interactions of Rydberg atoms and can be implemented at much larger interatomic distances than with dipole blockade. It also allows improving the individual addressing in the atomic quantum register by focused laser radiation.

Three-body F\"orster resonances were first demonstrated experimentally by us in Ref. \cite{17} for large disordered ensembles of Rydberg atoms of Cs, and in Ref.~\cite{18} the three-body nature of the interactions was confirmed for small mesoscopic ensembles containing $2-5$ Rb atoms. We also note that many-body electrically controlled F\"orster resonances for large ensembles of Rydberg atoms were studied experimentally and theoretically in Refs. \cite{19,20,21}, where the possibility of realizing four-body and higher resonances was discussed.

In Ref. \cite{22}, we proposed and analyzed a new type of three-body F\"orster resonance $3\times nP_{3/2} \to nS_{1/2} +(n+1)S_{1/2} +nP_{1/2} $, which can be realized with Rb Rydberg atoms for an arbitrary principal quantum number $n$. This resonance will be further named fine-structure-state-changing (FSSC) resonance \cite{15}. Its peculiarity is that the third atom goes into a state with a total moment $J=1/2$, which has no Stark structure, so two-body F\"orster resonances are completely absent. This distinguishes it from the usual three-body resonance $3\times nP_{3/2} \rightarrow nS_{1/2}+(n+1)S_{1/2}+nP_{3/2}^{\star}$, where the third atom changes only the moment projection $M$. One of the drawbacks of the latter resonance is the proximity of the two-body F\"orster resonance $2\times nP_{3/2} \rightarrow nS_{1/2}+(n+1)S_{1/2}$, which partially overlaps with the three-body resonance in the electric field scale \cite{18}.

In the subsequent theoretical work, we proposed a scheme for implementing the three-qubit Toffoli quantum gate based on FSSC three-body F\"orster resonances \cite{15}. Also, a scheme for implementing doubly controlled phase gates CC$\Phi $ based on these resonances with the addition of a radio-frequency field creating additional Rydberg Floquet levels was developed \cite{16}. We have studied two-body F\"orster resonances for Rydberg Floquet levels both experimentally and theoretically in Refs. \cite{23,24}.

In our recent theoretical paper \cite{25}, an extended theoretical study of the FSSC three-body F\"orster resonance was performed for various spatial configurations of three interacting Rb Rydberg atoms and conditions for their experimental implementation were determined. It was found that in a linear spatial configuration of three atoms, the three-body resonance splits into two resonances. In this case, one of the resonances has a weaker dependence of the resonant electric field on the distance between the atoms and is therefore the most suitable for performing experiments on observing coherent population oscillations of collective three-body states and implementing three-qubit quantum gates based on them.

In this paper, the splitting and shifts of the FSSC three-body F\"orster resonances are investigated analytically in more detail to identify their physical mechanisms and possible applications for probing the three-body interactions between single Rydberg atoms. These issues have not been studied previously. In particular, the analytical model presented in Ref. \cite{25} concerned the triangular configuration and did not properly analyze the splitting.

\section{Analytical model}

\begin{figure}
\includegraphics[scale=0.8]{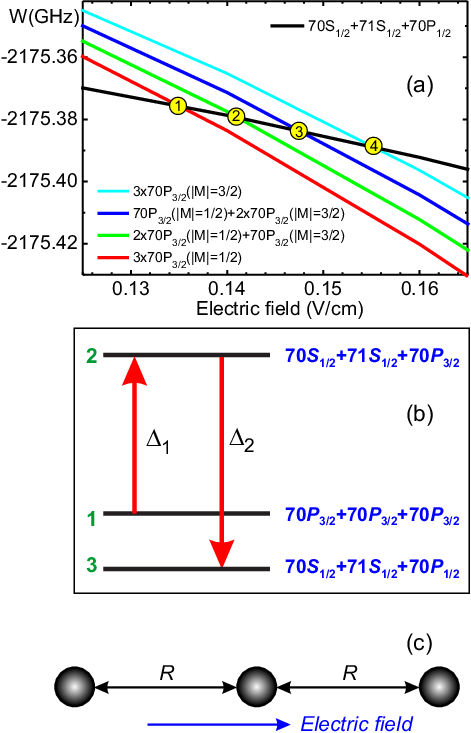}
\caption{\label{Fig1} (a) Numerically calculated Stark structure of the FSSC three-body F\"orster resonance $3\times 70P_{3/2} \to 70S_{1/2} +71S_{1/2} +70P_{1/2} $ for three Rb Rydberg atoms. The energies W of various three-body collective states are shown versus the controlling dc electric field. Intersections between collective states (labeled by numbers) correspond to the F\"orster resonances. (b) Simplified scheme of the FSSC three-body F\"orster resonance $3\times 70P_{3/2} \to 70S_{1/2} +71S_{1/2} +70P_{1/2} $ for three Rb Rydberg atoms. The initially populated collective state is state 1. The intermediate collective state is state 2, with one atom remaining in the initial $70P_{3/2} $ state. The final collective state is state 3, with the changed fine-structure component of the $P$ state. The energy defects $\Delta _{1} $ and $\Delta _{2} $ are controlled by the dc electric field. The three-body F\"orster resonance occurs at $\Delta _{1} \approx \Delta _{2} $. (c)~The linear spatial configuration of the three Rydberg atoms considered in this paper.}
\end{figure}

Figure~\ref{Fig1}(a) presents the numerically calculated Stark structure of the FSSC three-body F\"orster resonance $3\times 70P_{3/2} \to 70S_{1/2} +71S_{1/2} +70P_{1/2} $ for three Rb Rydberg atoms, which was first considered in Ref. \cite{22}. The energies W of various three-body collective states are shown versus the controlling dc electric field. The intersections between collective states (labeled by numbers) correspond to the F\"orster resonances. 

For three Rydberg atoms in the initial state $70 P_{3/2}(|M| =1/2)$, the three-body F\"orster resonance 1 in Fig.~\ref{Fig1}(a) corresponds to the resonant transition between collective states $3\times 70P_{3/2}(|M|=1/2) \to 70S_{1/2} +71S_{1/2} +70P_{1/2} $. This transition is, in fact, composed of the two nonresonant two-body relay transitions $3\times70P_{3/2}(|M| =1/2) \rightarrow 70S_{1/2}+71S_{1/2}+70P_{3/2}(|M|=1/2) \rightarrow 70S_{1/2}+71S_{1/2}+70P_{1/2}$ occurring simultaneously. The latter occurs due to the nonresonant exchange interactions $70P_{3/2}+70S \rightarrow 70S+70P_{1/2}$ or $70P_{3/2}+71S \rightarrow 71S+70P_{1/2}$ corresponding to the excitation hopping between $S$ and $P$ Rydberg atoms \cite{17,26}. Despite the use of a relay, the transfer occurs in a single step, implying a Borromean character of the relay atom, which absorbs the energy of the finite F\"orster defect.

Figure~\ref{Fig1}(b) presents the simplified scheme of the FSSC three-body F\"orster resonance $3\times 70P_{3/2} \to 70S_{1/2} +71S_{1/2} +70P_{1/2} $ for three Rb Rydberg atoms. The initially populated collective state 1 is $3\times 70P_{3/2} $. The final collective state 3 is $70S_{1/2} +71S_{1/2} +70P_{1/2} $, with the changed fine-structure component of the $P$ state. The intermediate collective state 2 is $70S_{1/2} +71S_{1/2} +70P_{3/2} $. The energy defects $\Delta _{1} $ and $\Delta _{2} $ are controlled by the dc electric field. The value of $\Delta _{1} $ can be varied significantly, while $\Delta _{2} $ is nearly constant in the vicinity of the F\"orster resonance, which occurs at $\Delta _{1} \approx \Delta _{2} $.

\begin{figure*}
\includegraphics[scale=0.9]{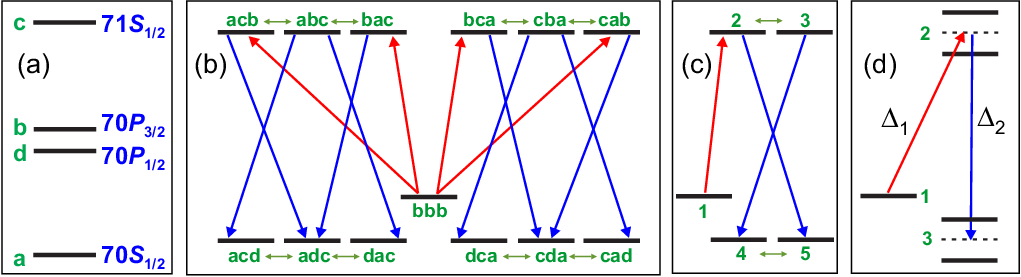}
\caption{\label{Fig2} (a) Rydberg states (labeled $a-d$) in a single Rb atom related to the FSSC three-body F\"orster resonance $3\times 70P_{3/2} \to 70S_{1/2} +71S_{1/2} +70P_{1/2} $. (b) Collective states of the three interacting Rb Rydberg atoms. Their labels $ijk$ indicate the related states of Fig.~\ref{Fig2}(a) and take into account all possible atom permutations. Red arrows indicate interaction-induced transitions from the initial state $bbb$ ($3\times 70P_{3/2}$) to the intermediate states of the kind $70S_{1/2} +71S_{1/2} +70P_{3/2} $. Blue arrows indicate interaction-induced transitions from the intermediate states to the final states of the kind $70S_{1/2} +71S_{1/2} +70P_{1/2} $. Green horizontal arrows indicate always-resonant exchange transitions corresponding to the excitation hopping between $S$ and $P$ Rydberg atoms. (c) Reduced scheme of the three-body F\"orster resonance that takes into account the symmetries and identities of some transitions in Fig.~\ref{Fig2}(b). State 1 is the same as state $bbb$. State 2 represents identical states $acb, bac, bca, cab$. State 3 represents identical states $abc, cba$. State 4 represents identical states $acd, dac, dca, cad$. State 5 represents identical states $adc, cda$. States 2-3 and 4-5 still experience always-resonant exchange transitions, which should result in their mixing and dynamic splitting. (d)~Final reduced scheme of the three-body F\"orster resonance. States 2 and 3 of Fig.~\ref{Fig1}(c) are replaced by two split states labeled as 2. States 4 and 5 of Fig.~\ref{Fig1}(c) are also replaced by two split states labeled as 3. As the splittings due to always-resonant dipole-dipole interaction are significant, in the analytical calculations we can take into account only one of the two states in each mixed state 2 or 3. }
\end{figure*}

Three-body F\"orster resonances are not described by the two-body operator of dipole-dipole interaction. This requires a special theoretical model to be developed. It is a rather complicated problem, since we should take into account all Stark and magnetic sublevels of the interacting Rydberg atoms. Therefore, we will consider a simplified analytical model for three frozen Rydberg atoms. We have first derived a similar model for the three-body F\"orster resonance ${\rm 3}\times nP_{3/2} \to nS_{1/2} +(n+1)S_{1/2} +nP_{3/2}^{\star} $ in Ref. \cite{27} for an equilateral triangle configuration, when the interaction energy for each atom pair was equal. 

However, our numerical simulations in Ref. \cite{27} have shown that the triangle configuration actually delivers many three-body interaction channels, which cannot be resolved in the electric field scale. This was also true for the FSSC F\"orster resonance ${\rm 3}\times nP_{3/2} \to nS_{1/2} +(n+1)S_{1/2} +nP_{1/2} $ \cite{22}. Both papers \cite{22,27} have found that optimal spatial configuration of the three interacting Rydberg atoms is a linear chain with interatomic distance $R$ aligned along the dc electric field, as depicted in Fig.~\ref{Fig1}(c). Due to specific selection rules, which are discussed below, there are only two interaction channels that deliver only two well-resolved three-body F\"orster resonances in the electric field scale. Therefore, in this paper we will consider only this linear spatial configuration. We can also note that the side atoms in Fig.~\ref{Fig1}(c) interact mainly with the central atom, so the interactions between the side atoms can be neglected in the analytical calculations, since these are 8 or 64 times smaller for the resonant dipole-dipole or nonresonant van der Waals interactions.

Figure~\ref{Fig2}(a) shows Rydberg states in a single Rb atom related to the considered three-body F\"orster resonance $3\times 70P_{3/2} \to 70S_{1/2} +71S_{1/2} +70P_{1/2} $. State ${\left| 70S_{1/2} \left(M=1/2\right) \right\rangle} $ is labeled as $a$,  state ${\left| 70P_{3/2} \left(M=1/2\right) \right\rangle} $ is labeled as $b$,  state ${\left| 71S_{1/2} \left(M=1/2\right) \right\rangle} $ is labeled as $c$,  and state ${\left| 70P_{1/2} \left(M=1/2\right) \right\rangle} $ is labeled as $d$.  

For the geometry and quantization axis of Fig.~\ref{Fig1}(c), only interaction-induced transitions that do not change the total  moment projection $M$ are allowed. The calculated $z$ components of the matrix elements of dipole moments of allowed transitions between the above states are $d_{ba} =2395$ a.u., $d_{bc} =2335$~a.u., $d_{da} =1721$ a.u., and $d_{dc} =1622$ a.u.. 

Figure~\ref{Fig2}(b) shows collective states of the three interacting Rb Rydberg atoms. Their labels $ijk$ indicate the related states of Fig.~\ref{Fig2}(a) and take into account all possible atom permutations. Red arrows indicate interaction-induced transitions from the initial state $bbb$ ($3\times 70P_{3/2} $) to the intermediate states of the kind $70S_{1/2} +71S_{1/2} +70P_{3/2} $. Blue arrows indicate interaction-induced transitions from the intermediate states to the final states of the kind $70S_{1/2} +71S_{1/2} +70P_{1/2} $. Green horizontal arrows indicate always-resonant exchange transitions corresponding to the excitation hopping between $S$ and $P$ Rydberg atoms. 

The time dynamics and line shapes of the three-body F\"orster resonances can be calculated using the Schr\"odinger equation for the amplitudes of all 13 collective states in Fig.~\ref{Fig2}(b). However, clearly understandable analytical formulas can be obtained only for a three-level system, as we have found in Ref. \cite{27}. Therefore, one needs to reduce Fig.~\ref{Fig2}(b) to an effective three-level system.

First, we can note that there are symmetries and identities of some transitions in Fig.~\ref{Fig2}(b), so we can finally obtain a reduced five-level scheme of the three-body F\"orster resonance shown in Fig.~\ref{Fig2}(c). Here, state 1 is the same as state $bbb$. State 2 represents identical states $acb$, $bac$, $bca$, $cab$. State 3 represents identical states $abc$, $cba$. State 4 represents identical states $acd$, $dac$, $dca$, $cad$. State 5 represents identical states $adc$, $cda$. States 2-3 and 4-5 still experience always-resonant exchange transitions, which should result in their mixing and dynamic splitting. Taking into account the permutation degeneracies of states 2-5, the amplitudes of collective states in Fig.~\ref{Fig2}(c) are described by the following equations, obtained from the Schr\"odinger equation:

\begin{equation} \label{Eq1} 
\begin{array}{l} {i\dot{a}_{1} =4\Omega _{12} a_{2} {\rm e}^{-i\Delta _{1} t} }, \\ {i\dot{a}_{2} =\Omega _{23} a_{3} +\Omega _{12} a_{1} {\rm e}^{i\Delta _{1} t} +\Omega _{25} a_{5} {\rm e}^{i\Delta _{2} t} }, \\ {i\dot{a}_{3} =2\Omega _{23} a_{2} +2\Omega _{34} a_{4} {\rm e}^{i\Delta _{2} t} }, \\ {i\dot{a}_{4} =\Omega _{45} a_{5} +\Omega _{34} a_{3} {\rm e}^{-i\Delta _{2} t} }, \\ {i\dot{a}_{5} =2\Omega _{45} a_{4} +2\Omega _{25} a_{2} {\rm e}^{-i\Delta _{2} t} }. \end{array} 
\end{equation} 

\noindent Here $\Omega _{ij} =V_{ij} /\hbar $ are matrix elements of dipole-dipole interactions with energies $V_{ij} $ in circular frequency units. The terms without exponents on the right-hand sides are responsible for the always-resonant exchange interactions that split the degenerate states 2-3 and 4-5, while the terms with the exponents drive the transitions between nondegenerate collective states. The dipole-dipole matrix elements $\Omega _{ij} $ are given by

\begin{equation} \label{Eq2} 
V_{ij} =\frac{d_{i} d_{j} }{4\pi \varepsilon _{0} } \left[\frac{1}{R^{3} } -\frac{3\, \, Z^{2} }{R^{5} } \right],   
\end{equation} 

\noindent where $d_{i} ,\; d_{j} $ are dipole moments of transitions in a single atom, $Z$ is the $z$ component of the vector connecting the two atoms $R$ (the $z$ axis is chosen along the dc electric field), and $\varepsilon_0$ is the dielectric constant.

Taking into account the fourfold level degeneracy of states 2 and 4, and twofold level degeneracy of states 3 and 5, which accounts for the atom permutations in Fig.~2(b), the three-atom resonance spectrum is then calculated as

\begin{equation} \label{Eq3} 
\rho _{3} =\frac{4}{3}(|a_{2}|^{2} +|a_{4}|^{2})+\frac{2}{3}(|a_{3}|^{2} +|a_{5}|^{2}). 
\end{equation} 

\noindent This value corresponds to the probability to find one of the three atoms in the final $70S_{1/2} $ state and it is the signal measured in our experiments \cite{18}.

Second, Eqs. (\ref{Eq1}) cannot be solved analytically, yet, for the arbitrary interaction energy, detunings, and time $t$. In order to reduce it to a three-level system, we can note that energies $\Omega _{23} $ and $\Omega _{45} $ of the always-resonant exchange interactions in Eqs.~(\ref{Eq1}) are directly given by Eq.~(\ref{Eq2}), so these are rather strong ($\approx10$ MHz at $R=10 $~$\mu$m) and scale as $R^{-3}$. If we set the other interactions in Eqs.~(\ref{Eq1}) to be zero, states 2-3 and 4-5 turn out to be mixed and symmetrically split by $2\sqrt{2} \Omega _{23} $ and $2\sqrt{2} \Omega _{45} $, as shown in Fig.~\ref{Fig2}(d). These splittings are much stronger than nonresonant interactions $\Omega _{12} $, $\Omega _{25} $, and $\Omega _{34} $. The latter have large energy defects $\Delta _{1} $, $\Delta _{2} \approx 274$ MHz, and therefore are described by the weak second-order perturbation terms like $\Omega _{ij}^{2} /\Delta _{i} $ that scale as $R^{-6}$. 

With the above considerations in mind, the five-level system of Fig.~\ref{Fig2}(c) can be reduced to an effective three-level system of Fig.~\ref{Fig2}(d). Here, state 1 remains to be an initial state, intermediate state 2 is strongly split by always-resonant exchange interaction $2\sqrt{2} \Omega _{23} $ to two sublevels, and final state 3 is also strongly split by always-resonant exchange interaction $2\sqrt{2} \Omega _{45} $ to two sublevels. The splitting of state 2 is not essential, since three-body transition $1\to 3$ is driven with the intermediate detunings $\Delta _{1} $ and $\Delta _{2} $ being much larger than $2\sqrt{2} \Omega _{23} $. Then the split state 2 can be viewed as a single state 2. We should only take into account the splitting $2\sqrt{2} \Omega _{45} $ of state 3 that will result in the dynamic splitting of three-body resonance, thus resembling the Autler-Townes effect in probe spectroscopy of three-level systems driven by strong radiation \cite{28}. As this splitting is much larger than the expected three-body interaction energy, the three-body transitions to the two split sublevels of state 3 can be calculated independently.

Finally, the two split three-body resonances of Fig.~\ref{Fig2}(d) can be described as two effective three-level systems with the following equations:

\begin{equation} \label{Eq4} 
\begin{array}{l} {i\dot{b}_{1} =4\Omega _{12} b_{2} {\rm e}^{-i\Delta _{1} t} }, \\ {i\dot{b}_{2} =\Omega _{12} b_{1} {\rm e}^{i\Delta _{1} t} +\Omega _{25} b_{3} {\rm e}^{i\Delta _{2}^{*} \; t} }, \\ {i\dot{b}_{3} =\Omega _{25} b_{2} {\rm e}^{-i\Delta _{2}^{*} \; t} }. \end{array} 
\end{equation} 

\noindent Here, $\Delta _{2} $ is replaced by an effective detuning $\Delta _{2}^{*} =\Delta _{2} \pm \sqrt{2} \Omega _{45} $ for the two split sublevels of the final state 3. 

Equations (\ref{Eq4}) can be solved analytically by proper substitutions, since they can be reduced to a single cubic equation whose roots are found by the known mathematical formulas, as we did in Ref. \cite{27}. However, in order to simplify the previously proposed description, in this work we apply the adiabatic approximation. 

\begin{figure*}
\includegraphics[scale=1.0]{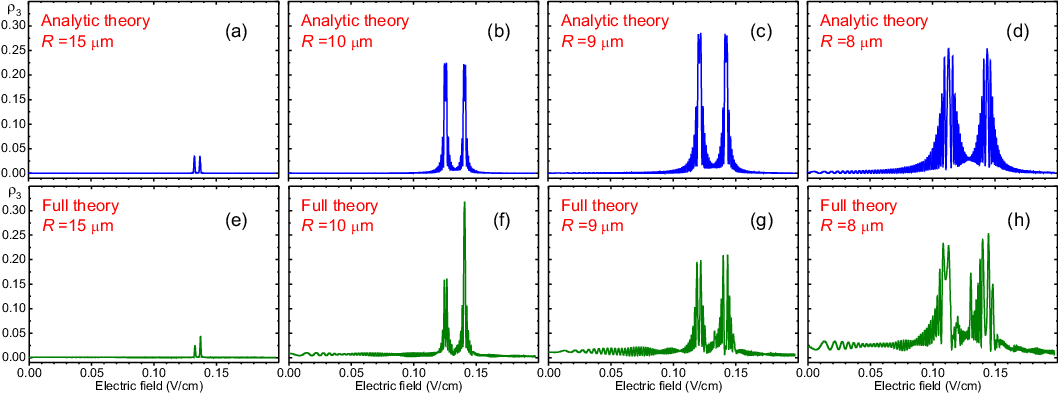}
\caption{\label{Fig3} Analytically (a)-(d) and numerically (e)-(h) calculated spectra of the FSSC three-body F\"orster resonance $3\times 70P_{3/2} \to 70S_{1/2} +71S_{1/2} +70P_{1/2} $ in Rb Rydberg atoms for the interaction time of 1 $\mu$s and various interatomic distances $R=$15, 10, 9, and 8~$\mu$m. Analytical calculations have been done with Eqs.~(\ref{Eq6}) and (\ref{Eq8}). Numerical calculations have been done with the full theoretical model developed by us earlier in Refs. \cite{14,27}. A good agreement in positions and heights of the split resonances is observed, thus justifying the validity of Eqs.~(\ref{Eq6}) and (\ref{Eq8}) to be used in measuring Rydberg interaction strength.}
\end{figure*}

We can note that, due to large intermediate detunings, intermediate state 2 in Fig.~\ref{Fig2}(d) is almost unpopulated at the three-body F\"orster resonance \cite{23}. Therefore this state can be adiabatically eliminated by replacing $b_{2} $ with a new variable whose rapidly oscillating part is extracted as $b_{2} =\beta _{2} e^{i\Delta_0 t} $. Here we are introducing an average intermediate detuning $\Delta _{0} =(\Delta _{1} +\Delta _{2}^{*} )/2$. Substituting $b_{2} $ in Eqs.~(\ref{Eq4}) and neglecting the small term with $\dot{\beta }_{2} $, we finally obtain the equations for an effective two-level system:

\begin{equation} \label{Eq5} 
\begin{array}{l} {i\dot{b}_{1} \approx -\displaystyle \frac{4\Omega _{12}^{2} }{\Delta _{0} } b_{1} -\frac{4\Omega _{12} \Omega _{25} }{\Delta _{0} } b_{3} {\rm e}^{-i\Delta t} }, \\ \\{i\dot{b}_{3} \approx -\displaystyle \frac{\Omega _{25}^{2} }{\Delta _{0} } b_{3} -\frac{2\Omega _{12} \Omega _{25} }{\Delta _{0} } b_{1} {\rm e}^{i\Delta t} }, \end{array}  
\end{equation} 

\noindent where $\Delta =\Delta _{1} -\Delta _{2}^{*} $. The terms without exponents on the right-hand sides are responsible for the dynamic shifts due to nonresonant Rydberg interactions, while the terms with the exponents drive the transitions between collective states 1 and 3 and induce coherent phase and population oscillations.

The analytical solution of Eqs.~(\ref{Eq5}) with the initial conditions $b_{1} (0)=1,\; b_{3} (0)=0$ and with an appropriately modified Eq.~(\ref{Eq3}) is given by the formula

\begin{equation} \label{Eq6} 
\rho _{3} \approx \frac{\Omega _{0}^{2} /3}{\left(\delta -\delta _{0} \right)^{2} +\Omega _{0}^{2} } \sin ^{2} \left[\frac{t}{2} \sqrt{\left(\delta -\delta _{0} \right)^{2} +\Omega _{0}^{2} } \right], 
\end{equation} 

\noindent where $\delta =\Delta _{1} -\Delta _{2} $ is the detuning from the unperturbed three-body resonance, $\delta _{0} =\pm \sqrt{2} \Omega _{45} +(\Omega _{25}^{2} -4\Omega _{12}^{2} )/\Delta _{0} $ is the interaction-induced splitting and shift of the three-body resonance, and $\Omega _{0} =4\Omega _{12} \Omega _{25} /\Delta _{0} $ is the Rabi-like population oscillation frequency. 

Compared to the analytical solution that we obtained in Ref. \cite{27} for an equilateral triangle configuration, the new feature is that in the linear spatial configuration along the $z$ axis [see Fig.~\ref{Fig1}(c)] the three-body resonance is split to two resonances, that occur in different electric fields at $\delta _{+} =\sqrt{2} \Omega _{45} +(\Omega _{25}^{2} -4\Omega _{12}^{2} )/\Delta _{0} $ and $\delta _{-} =-\sqrt{2} \Omega _{45} +(\Omega _{25}^{2} -4\Omega _{12}^{2} )/\Delta _{0} $. Therefore, Eq.~(\ref{Eq6}) in fact should be replaced by the sum of two such solutions, one at $\delta _{+}$ and another at $\delta _{-}$. They have different dependences on the interatomic distance $R$, which we analyze below. In what follows we will calculate the numerical values of the splitting, shift and Rabi frequency in their dependences on $R$ for possible usage in measurements of the Rydberg interaction strength, which we propose in this paper.

First, we note that the radial parts of dipole moments between relevant $S$ and $P$ Rydberg states are nearly identical with the average value of 5017~a.u.. Then the values of $\Omega _{12} $, $\Omega _{25} $, and $\Omega _{45} $ are related via angular parts of dipole moments as $\Omega _{12} =2\Omega /9$, $\Omega _{25} =\sqrt{2} \Omega /9$, and $\Omega _{45} =\Omega /9$, where $\Omega =4.9\times 10^4 R^{-3}$ (MHz) and $R$ is taken in micrometers.

The three-body detuning $\delta =\Delta _{1} -\Delta _{2} $ and intermediate detuning $\Delta _{0} =(\Delta _{1} +\Delta _{2}^{} )/2$, calculated from Fig.~\ref{Fig1}(a), have the following dependences on the electric field $F$ (taken in V/cm units) near the resonance:

\begin{equation} \label{Eq7} 
\begin{array}{l} {\delta =-72.51+53.5F+3586\; F^{2} \quad ({\rm MHz})}, \\ \\ {\Delta _{0} =248.51-14.84F+1518\; F^{2} \quad ({\rm MHz})}. \end{array} 
\end{equation} 

\noindent The three-body detuning becomes zero at $F\approx 0.135$~V/cm. In this field, the intermediate detuning is $\Delta _{0} \approx 274$~MHz, and it remains nearly constant as $F$ is scanned across the three-body resonance. Then the numerical formulas for the positions and Rabi frequency of the two resonances are given by

\begin{equation} \label{Eq8} 
\begin{array}{l} {\delta _{+} =\displaystyle \frac{7698}{R^{3} } -\displaystyle \frac{1.52\times 10^{6} }{R^{6} } \quad ({\rm MHz})}, \\ \\{\delta _{-} =-\displaystyle \frac{7698}{R^{3} } -\displaystyle \frac{1.52\times 10^{6} }{R^{6} } \quad ({\rm MHz})}, \\ \\{\Omega _{0} =\displaystyle \frac{1.224\times 10^{6} }{R^{6} } \quad ({\rm MHz})}, \end{array} 
\end{equation} 

\noindent where $R$ is taken in micrometers.

\section{Comparison with numerical simulations in full theory}

Using Eqs.~(\ref{Eq6}) and (\ref{Eq8}), Figs.~\ref{Fig3}(a)--\ref{Fig3}(d) present analytically calculated spectra of the FSSC three-body F\"orster resonance $3\times 70P_{3/2} \to 70S_{1/2} +71S_{1/2} +70P_{1/2} $ in Rb Rydberg atoms for the interaction time of 1 $\mu$s and various interatomic distances $R=$15, 10, 9 and 8~$\mu$m. As expected, the resonance is split into two resonances, and the splitting strongly depends on $R$. The resonance height grows as $R$ decreases and saturates at $R=9$ $\mu$m, where it starts to broaden and the Rabi-like population oscillations become visible in the resonance wings.

\begin{figure}
\includegraphics[scale=1.0]{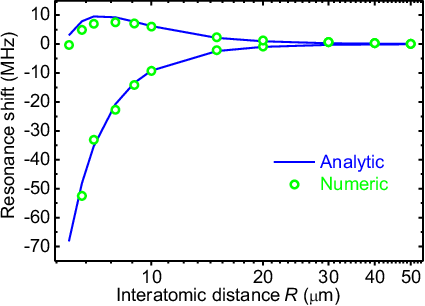}
\caption{\label{Fig4} Dependences of the shifts of the centers of the two split FSSC three-body resonances  $3\times 70P_{3/2} \to 70S_{1/2} +71S_{1/2} +70P_{1/2} $ in Fig.~\ref{Fig3} on the interatomic distance $R$ for the analytical (blue curves) and numerical (green circles) theoretical models. The shifts are recalculated from V/cm to the MHz scale using Eqs.~(\ref{Eq7}).}
\end{figure}

In order to check for the validity of simple equations~(\ref{Eq6}) and (\ref{Eq8}), we have also done the numerical calculations with the full theoretical model that we developed earlier in Refs.~\cite{14,27}. Numerical calculations of probability amplitudes of all collective states were performed based on the Schr\"odinger equation. A complete model of atomic interaction was used taking into account the Zeeman structure of Rydberg levels. To simplify calculations, collective states of the atomic system with an energy defect of more than 2 GHz in zero electric field were excluded from consideration. Thus, for the three-body F\"orster resonance in atoms in the initial state $70P_{3/2}(M=1/2)$, a complete calculation required taking into account 360 collective states with all possible values of the angular moment projections. Each collective state was a product of the states of three atoms in the basis states $70S_{1/2}(M = \pm1/2)$, $71S_{1/2}(M = \pm1/2)$), $70P_{1/2}(M = \pm1/2)$, $70P_{3/2}(M = \pm 1/2,\pm 3/2)$.

The finite radiative lifetimes of all Rydberg states, calculated according to Ref. \cite{29}, taking into account the effect of surrounding blackbody radiation at T=300 K ($70S$ -- 152 $\mu $s$;$ $71S$ -- 156 $\mu $s$;$ $70P_{1/2}$ -- 189 $\mu $s$;$ $70P_{3/2}$ -- 191 $\mu $s), were also phenomenologically taken into account by introducing a weak depletion of the probability amplitudes into the Schr\"odinger equation. Although this leads to non-conservation of the total initial population of the collective states, this procedure allows us to calculate the maximum possible contrast of population oscillations for the implementation of three-qubit quantum gates \cite{14,15,16}.

Results of numerical simulations for the same interaction time of 1 $\mu$s are presented in Figs.~\ref{Fig3}(e)--\ref{Fig3}(h). The obtained numerical spectra well agree with the analytical ones in Figs.~\ref{Fig3}(a)--\ref{Fig3}(d), although analytical equations~(\ref{Eq6}) and (\ref{Eq8}) were obtained with numerous approximations and simplifications. Figure~\ref{Fig4} shows the dependences of the shifts of the centers of the two split three-body resonances in Fig.~\ref{Fig3} on the interatomic distance $R$ for the analytical (blue curves) and numerical (green circles) theoretical models. The shifts are recalculated from the V/cm to the MHz scale using Eqs. (\ref{Eq7}). The spectra demonstrate full agreement in the positions of the two split three-body resonances for any $R$. 

The resonance peak heights are also close for the analytical and numerical calculations in Fig.~\ref{Fig3}. The only discrepancy can be noted in the relative peak heights and phases of population oscillations. This can be attributed to somewhat different effective interaction energies that appear in the full model, as well as to the interactions of the side atoms accounted for in the full model but neglected in the analytical one. We have checked in additional numerical simulations that taking into account the side atoms interaction strengthens one resonance while weakens the other.

A good agreement in positions and heights of the split three-body resonances thus justifies the validity of our simple analytical model.

\section{Discussion}

Analytically obtained equations (\ref{Eq6}) and (\ref{Eq8}) reveal several important features behind the physics of the split three-body F\"orster resonances.

First, the first-order terms in Eqs.~(\ref{Eq8}), which are proportional to $R^{-3}$, are responsible for the splitting of the three-body F\"orster resonance due to always-resonant dipole-dipole interactions of the degenerate final collective states. At long distances ($R>10$ $\mu$m) the splitting is symmetrical with respect to the unperturbed three-body resonance at $F=0.135$ V/cm, as can be seen from Fig.~\ref{Fig4}. The splitting thus presents a direct measure of the resonant dipole-dipole interaction strength between neighboring Rydberg atoms. From the measured splitting we can determine both the interaction energy and the distance between the atoms using Eqs.~(\ref{Eq8}).

Second, the second-order terms in Eqs.~(\ref{Eq8}), which are proportional to $R^{-6}$, are responsible for the dynamic shift of the three-body F\"orster resonance due to nonresonant van der Waals interactions of Rydberg atoms in the intermediate states. This shift has the same sign for the two split F\"orster resonances. It becomes observable at short distances ($R<10$ $\mu$m), as can be seen from Fig.~\ref{Fig4}. Therefore, from the measured shift and splitting we can determine also the van der Waals interaction energy between neighboring Rydberg atoms using Eqs.~(\ref{Eq8}).

Third, one can notice in Eqs.~(\ref{Eq8}) and Fig.~\ref{Fig4} that the resonance at $\delta _{-} $ only drops in energy as $R$ decreases. In contrast, the resonance at $\delta _{+} $ first grows in energy due to the splitting; then it goes to the maximum energy at $R=7.3$~$\mu$m and starts to drop only at shorter $R$. Therefore, in the vicinity of  $R=7.3$~$\mu$m, this resonance has a smooth plateau where it is insensitive to small fluctuations of $R$, which are always present in experiments with single atoms in optical dipole trap arrays. This resonance is thus most suitable for performing experiments to observe coherent oscillations of populations of collective three-body states and implement three-qubit quantum gates based on them, as it was pointed out in our recent paper \cite{25}. In Ref. \cite{25} we intentionally introduced fully random (non-symmetric) spatial fluctuations of the atom positions in order to simulate what we can get in real experiments. We have found that resonance at $\delta _{+} $ was less affected by the fluctuations. It remained narrow and exhibited population oscillations even at $20\%-40\%$ atom position uncertainty. This means that we can really observe these oscillations in our future experiments.

A possible experiment on the considered three-body F\"orster resonances can be arranged with three single Rb atoms in three optical dipole traps formed by a spatial light modulator that provides arbitrary spatial configuration and distance between the atoms \cite{3}. Then the three atoms are coherently laser-excited to an initial Rydberg state $nP_{3/2}$ by $\pi$ laser pulses \cite{3} and start to interact. The interaction is controlled by a dc electric field using many-body F\"orster resonances \cite{18,19,20,21}. The interaction-induced transitions between collective states can be observed optically using state-selective de-excitation of Rydberg states by $\pi$ laser pulses and by detecting the fluorescence of ground-state atoms. The spatial fluctuations of atoms are determined by the atom temperature and trap depth, and typically are below 1 $\mu$m. Therefore, the observation of narrow three-body resonances should be experimentally feasible in such an arrangement.

Fourth, Eq.~(\ref{Eq6}) shows that coherent Rabi-like population oscillations take place at three-body F\"orster resonance. At the exact resonances ($\delta =\delta _{0} $), the Rabi-like oscillation frequency is $\Omega _{0} $, which depends on the interaction strength according to Eqs.~(\ref{Eq8}). The maximum height of the resonance is 1/3 (one of the three atoms is found to be in the final $70S_{1/2}$ state). The resonance saturates and broadens when the interaction strength increases. The resonance width is determined by a combination of the Fourier width of the interaction pulse and of the three-body interaction strength $\Omega _{0} $.

Finally, observation of the split three-body F\"orster resonance in the scheme of Fig.~\ref{Fig2}(d) demonstrates full analogy with the Autler-Townes effect in a three-level system \cite{28}, when strong laser radiation on one transition induces splitting of its energy levels due to ac Stark effect, while another weak radiation probes this splitting on an adjacent transition. Therefore, the three-body F\"orster resonance can serve as a probe to measure the dipole-dipole interaction strength in Rydberg-atom arrays for applications in quantum information.

We note that the Rydberg interaction strength can also be measured by observing coherent population oscillations, as it was demonstrated for two-body interactions with two Rydberg atoms \cite{30,31,32,33,34,35,36} and in atom ensembles \cite{37,38}. Such oscillations, however, are hard to observe experimentally due to atom position fluctuations and parasitic electric fields, which are always present in experiments. The splitting of three-body F\"orster resonance can be an alternative method that would work even when the population oscillations are not observable.

\section{Conclusions}

In this paper we theoretically investigated the structure of the fine-structure-state-changing three-body F\"orster resonances in a linear spatial configuration of the three interacting Rydberg atoms. We built a relatively simple analytical model and found the approximate formulas for the time dynamics, line shape, dynamic splitting, and shift of the F\"orster resonance. This model clearly reveals the physics behind the resonance structure. In particular, the splitting appears due to always-resonant dipole-dipole interaction of the degenerate final collective states, while the dynamic shift appears due to nonresonant van der Waals interactions of intermediate states. 

A comparison of the simple analytical model with a more precise numerical model, which takes into account Zeeman sublevels of all Rydberg states, has shown a very good agreement for the splitting and shifts, thus demonstrating the validity of the analytical model.

The splitting and shifts observed in experiments can serve as a probe of the dipole-dipole and van der Waals interaction strengths in Rydberg-atom arrays for applications in quantum information.

\begin{acknowledgments}
This work was supported by the Russian Science Foundation Grant No. 23-12-00067, https://rscf.ru/project/23-12-00067/. I.N.A. and P.C. were supported by the French National Research agency (ANR) under Grant No. ANR-22-CE47-0005 (QIPRYA project).
\end{acknowledgments}

\section*{DATA AVAILABILITY}
The data that support the findings of this article are available from the corresponding author upon reasonable request.

\end{document}